\documentclass[aps,prl,twocolumn,superscriptaddress]{revtex4}
\usepackage{longtable}
\usepackage{graphicx}
\usepackage{dcolumn}
\usepackage{bm}
\usepackage{amsmath}
\usepackage[psamsfonts]{amssymb}

\newcommand{\STO}{SrTiO$_3$}
\newcommand{\LAO}{LaAlO$_3$}

\newcommand{\etal}{\emph{et al.}}

\newcommand{\TiOtwo}{TiO$_2$}
\newcommand{\Rs}{$R_\Box$}

\begin{document}

\title{Solution Monolayer Epitaxy for Tunable Atomically Sharp Oxide Interfaces}


\author{A.  Ron}
\affiliation{Raymond and Beverly Sackler School of Physics and Astronomy, Tel-Aviv University, Tel Aviv, 69978, Israel}
\author{A. Hevroni}
\affiliation{Raymond and Beverly Sackler School of Chemistry, Tel-Aviv University, Tel Aviv, 69978, Israel}
\author{E. Maniv}
\affiliation{Raymond and Beverly Sackler School of Physics and Astronomy, Tel-Aviv University, Tel Aviv, 69978, Israel}
\author{M. Mograbi}
\affiliation{Raymond and Beverly Sackler School of Physics and Astronomy, Tel-Aviv University, Tel Aviv, 69978, Israel}
\author{L. Jin}
\affiliation{Peter Gr\"unberg Institute and Ernst Ruska-Centre for Microscopy and Spectroscopy with Electrons, Research Centre J\"ulich, D-52425 J\"ulich, Germany}
\author{C.-L. Jia}
\affiliation{Peter Gr\"unberg Institute and Ernst Ruska-Centre for Microscopy and Spectroscopy with Electrons, Research Centre J\"ulich, D-52425 J\"ulich, Germany}
\author{K. W. Urban}
\affiliation{Peter Gr\"unberg Institute and Ernst Ruska-Centre for Microscopy and Spectroscopy with Electrons, Research Centre J\"ulich, D-52425 J\"ulich, Germany}
\author{G. Markovich}
\affiliation{Raymond and Beverly Sackler School of Chemistry, Tel-Aviv University, Tel Aviv, 69978, Israel}
\author{Y. Dagan}
\affiliation{Raymond and Beverly Sackler School of Physics and Astronomy, Tel-Aviv University, Tel Aviv, 69978, Israel}
\email[]{yodagan@post.tau.ac.il}


\date{\today}

\begin{abstract}
Epitaxial growth of atomically-sharp interfaces serves as one of the main building blocks of nanofabrication. Such interfaces are crucial for the operation of various devices including transistors, photo-voltaic cells, and memory components. In order to avoid charge traps that may hamper the operation of such devices, it is critical for the layers to be atomically-sharp. Fabrication of atomically sharp interfaces normally requires ultra-high vacuum techniques and high substrate temperatures. We present here a new self-limiting wet chemical process for deposition of epitaxial layers from alkoxide precursors. This method is fast, cheap, and yields perfect interfaces as we validate by various analysis techniques. It allows the design of heterostructures with half-unit cell resolution. We demonstrate our method by designing hole-type oxide interfaces \STO/BaO/\LAO. We show that transport through this interface exhibits properties of mixed electron-hole contributions with hole mobility exceeding that of electrons. Our method and results are an important step forward towards a controllable design of a p-type oxide interface.
\end{abstract}

\maketitle


\begin{figure}
 \includegraphics[width=1\hsize]{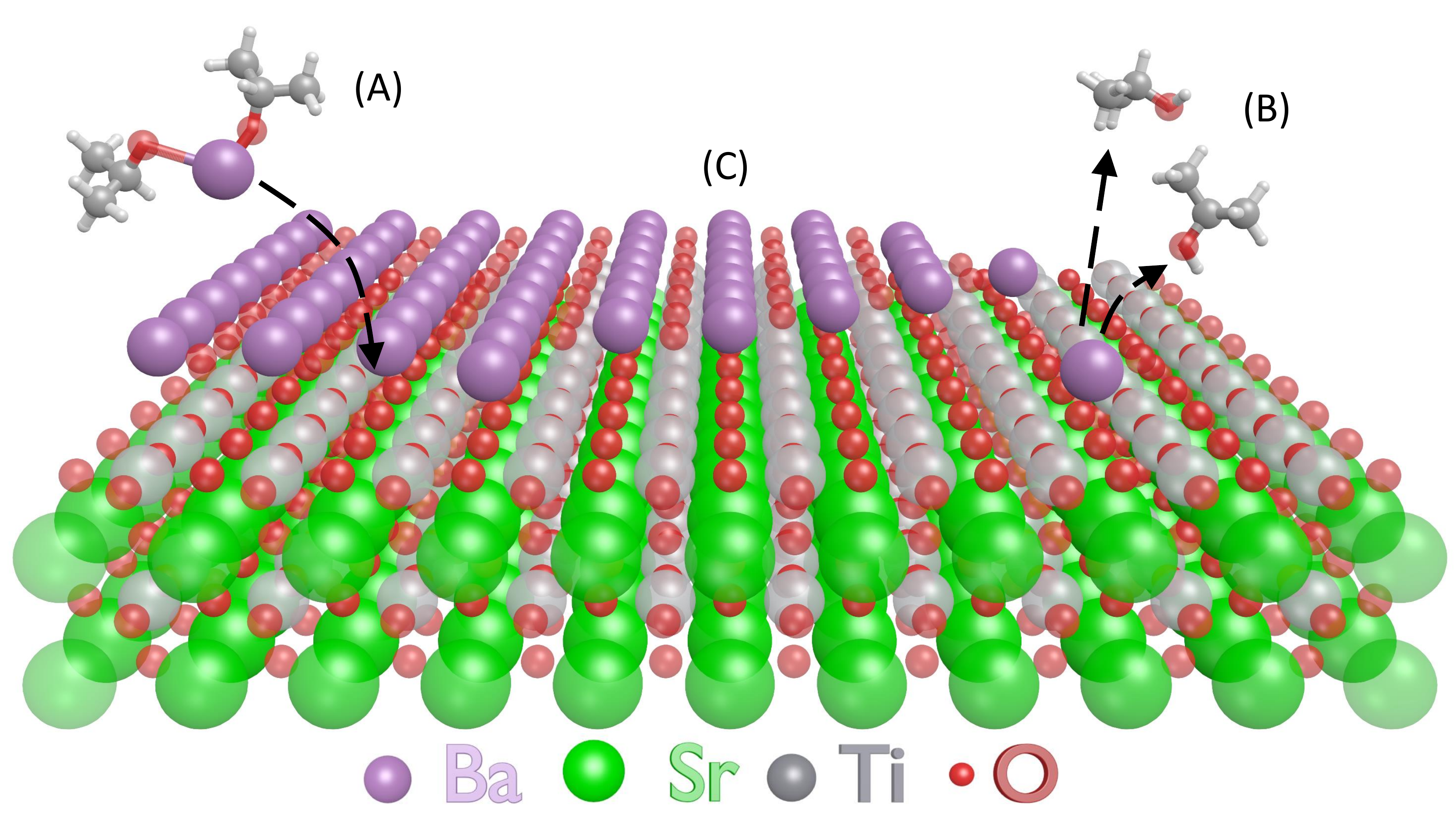}
  \caption {Solution monolayer epitaxy process. The TiO$_2$ terminated \STO~ is submersed in a solution of Ba $^{II}$ isopropoxide (A) in ethylene glycol, which is kept slightly below the decomposition temperature of the precursor  Ba $^{II}$ isopropoxide. Surface catalysis results in decomposition of of the precursor, which results in epitaxial deposition of BaO on the surface (B). This process is terminated when a complete layer of BaO covers the surface (C).
}\label{sketch}
\end{figure}
\begin{figure}
 \includegraphics[width=1\hsize]{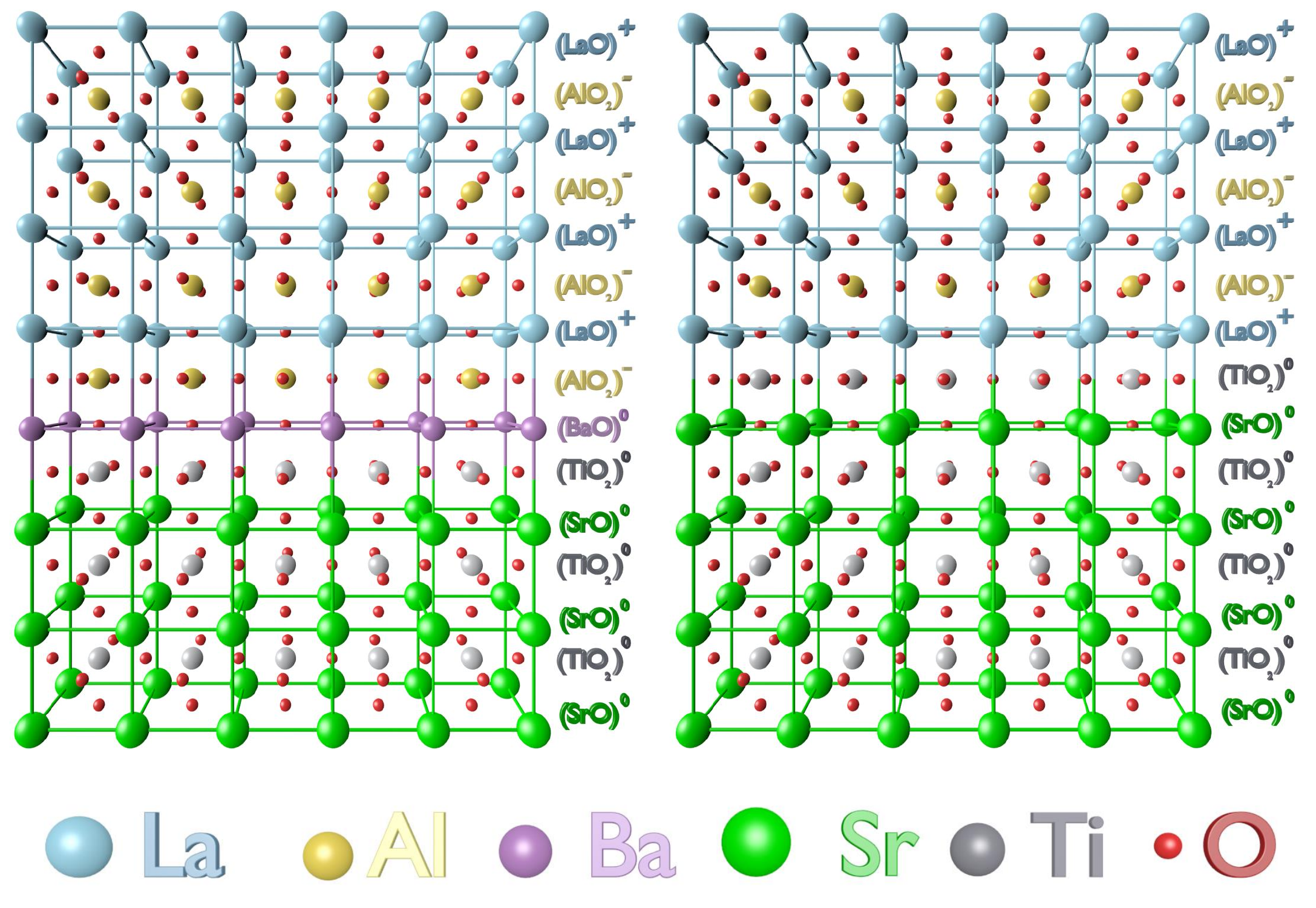}
  \caption {(color online). Two models of \STO/\LAO~  interfaces. On the right the conventional n-type interface. On the left our new p-type interface which incorporates a monolayer of BaO between the \STO~ substrate and the polar \LAO~ film. The presence of the Ba film forces a growth with inverted \LAO~ polarity.
}\label{Fig:1}
\end{figure}
\par
Growth methods of epitaxial thin films can be roughly categorized as physical and chemical. While physical methods (i.e. molecular beam epitaxy, pulsed laser deposition (PLD)\cite{smith1965vacuum}) are based on creating a beam of the film material and transporting it in vacuum onto the substrate. Chemical methods such as: chemical vapor deposition and atomic layer deposition (ALD) use a chemical precursor, a compound containing the film growth material. The precursor is transferred in vacuum onto the substrate, where the surface catalyzes the precursor dissociation reaction, and the deposition of the film. While giving very good results for film growth, these methods lack the versatility and become increasingly complex \cite{ma2008flex} when a wide variety of surface monolayers is required.
\par
In Solution monolayer epitaxy (SoME) the substrate of choice, in this case a (100) \TiOtwo~ terminated \STO, is submersed in a solution of a dissolved precursor of choice, at a temperature slightly lower than its decomposition temperature. Under these conditions the precursor molecules do not decompose in the solution unless they are in close proximity to the surface of the substrate, which catalyzes the decomposition of the precursor and the required material resulting in a monolayer.
\par
In our case SoME is used to grow a BaO monolayer (termination) on a \STO~ substrate as described in Figure \ref{sketch}. We then grow an additional epitaxial layer of \LAO~ using traditional pulsed laser deposition, giving us a \STO/BaO/\LAO~ structure as depicted in Figure \ref{Fig:1}(a) for comparison we show the standard \STO/\LAO~ interface in Figure \ref{Fig:1}(b). More details of our process are found in the methods section.
\par
Oxide interfaces can combine degrees of freedom from their constituents with emergent phenomena such as, superconductivity \cite{reyren2007superconducting, caviglia2008electric} magnetism \cite{shalom2009anisotropic, bert2011direct, dikin2011coexistence, salman2012nature, lee2013titanium, ron2014anomalous} tunable spin-orbit interaction \cite{shalom2010tuning, caviglia2010tunable} and quantum transport \cite{tsukazaki2007quantum, ron2014one, cheng2015electron}. The appearance of conductivity in the hallmark interface between \LAO~ and \STO~ has been explained as a result of stacking polar layers of \LAO~ thus accumulating potential energy \cite{ohtomo2004high, okamoto2004electronic, pentcheva2009avoiding,coey2013origin, popovic2008origin}. According to these scenario one expects the two interfaces presented in Figure \ref{Fig:1} to have carriers of opposite signs as explained in this figure.
\par
\begin{figure}[!ht]
\centering
 \includegraphics[width=1\hsize]{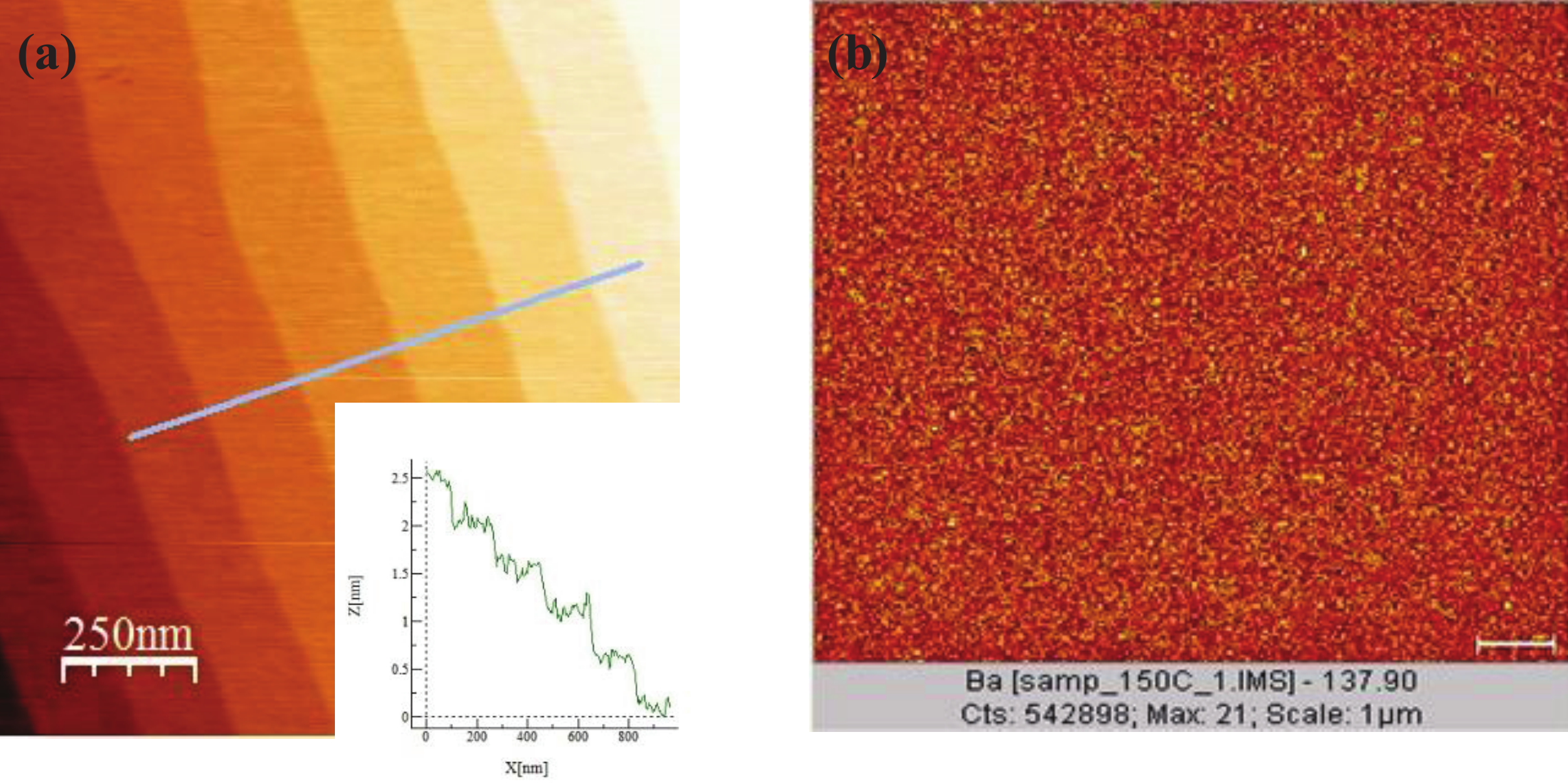}
  \caption {(a) AFM topography image of the surface after the SoME deposition of Ba, the terraces are 1uc high and are singly terminated. (b) TOF-SIMS element mapping of Ba. The coverage of Ba is uniform across the sample.
}\label{Fig:2}
\end{figure}

Prior to the \LAO~ deposition the SoME BaO deposited layer was characterized using an atomic force microscope (AFM) and time of flight secondary ion mass spectroscopy (TOF-SIMS), see Figure \ref{Fig:2}. Panel (a) shows an AFM topography image of the surface, a smooth surface is observed indicating full coverage of the substrate. Panel (b) shows a spatial distribution map of Ba as measured by TOF-SIMS showing a uniform distribution of Ba on the surface. The combination of these two techniques shows that we have indeed created a smooth single terminated BaO layer on top of the \STO~ substrate.

\par
\begin{figure}
 \includegraphics[width=1\hsize]{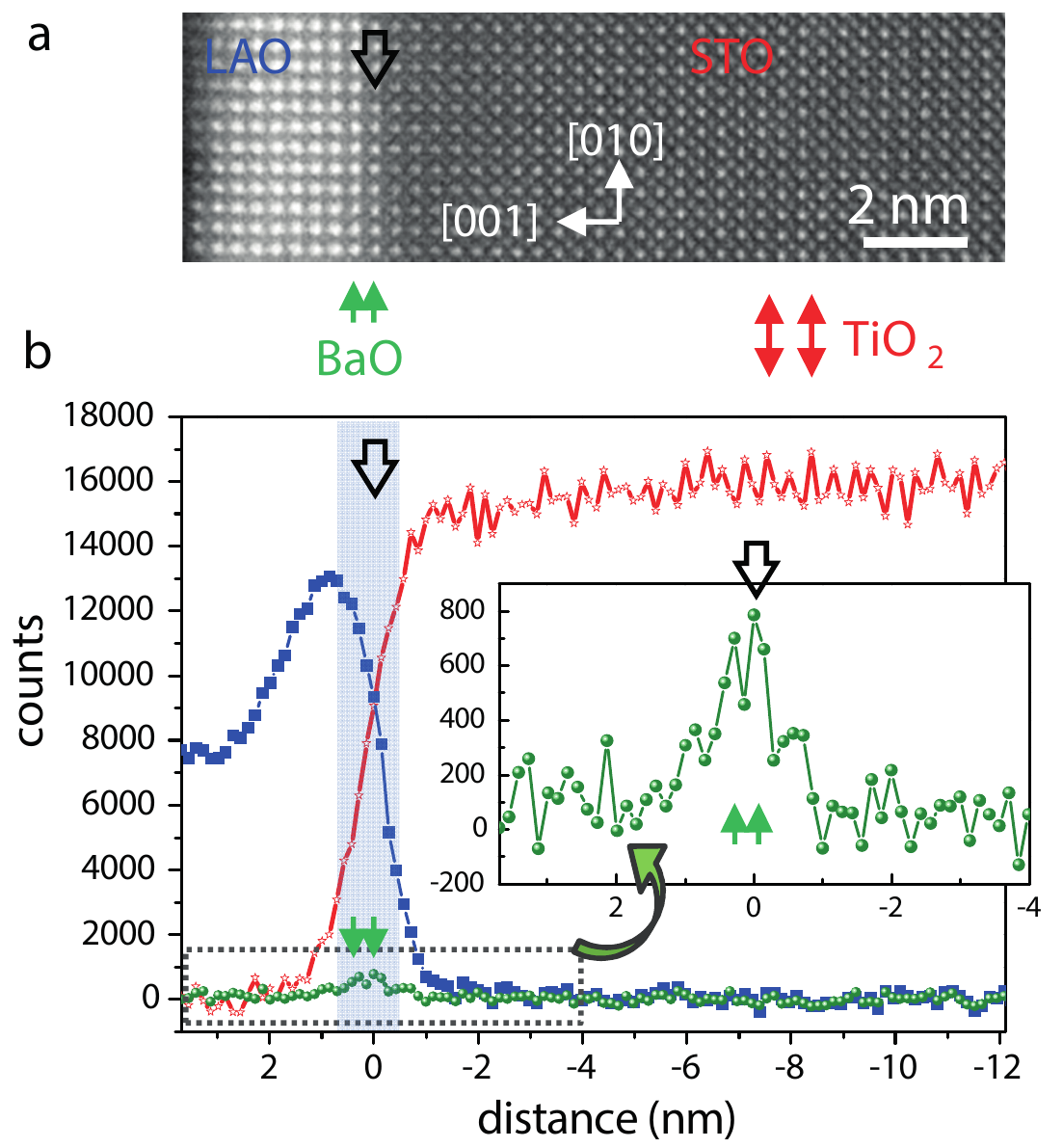}
  \caption {Atomic-resolution $Z$-contrast image and the elemental dispersion of the \LAO/BaO/\STO~ heterostructure. The measurements were made along the [100] crystallographic direction of \STO~ substrate. (a) high-angle annular dark-field image recorded before the EELS measurement showing the contrast difference between \LAO/BaO and \STO. The interface between \LAO/BaO and \STO~ substrate was marked by the black arrow. (b) intensity line profiles of the Ti $L_3$ (red), Ba $M_4$, (green), and La $M_5$ (blue) edges, plotted as a function of the distance away from the interface (marked by the black arrow), showing the elemental dispersions along the [001] growth direction of the \LAO/BaO/\STO~ heterostructure. The profiles were obtained by using the integrated intensity from the energy-loss window ranging from 464.8 to 466.8 eV for Ti, from 801 to 804.2 eV for Ba, and from 834.4 to 836 eV for La, respectively. Intensity modulations of Ti are evident; the peaks represent the real stacking of \TiOtwo~ planes as indicated by the red double-headed arrows. The maximum interdiffusion width, shadowed in light blue, was estimated to be 3 perovskite unit cells. Inset of (b), magnification of the intensity profile showing the modulation of Ba signal (denoted by green arrows) from the region indicated by the dotted rectangle.}\label{Fig:3}
\end{figure}

Figure \ref{Fig:3}(a) shows an atomic-resolution high-angle annular dark-field (HAADF) image of the \STO/BaO/\LAO~ heterostructure taken from the [100] direction of \STO, whose contrast is roughly proportional to the square of the mean atomic number $Z$ in the atomic columns\cite{pennycook2000z}. By examining the image contrast, the interface between \STO~ and BaO/\LAO~ was identified and marked by the black arrow. The dark contrast on the right side of the interface corresponds to the \STO~ substrate since it possesses much smaller $Z$ numbers with respect to those of BaO and \LAO. In the left area, the \LAO~ and BaO show very bright contrast. However, distinguishing BaO from \LAO~ in the HAADF image is essentially impossible because the atomic numbers ($Z_{Ba} = 56$, $Z_{La} = 57$) are almost the same.
\par
Figure \ref{Fig:3}(b) displays the intensity line profiles of Ti $L_2$, Ba $M_4$, and La $L_5$ edges extracted from the background subtracted electron energy-loss spectroscopy (EELS) dataset (supplemental Figure S1(b)), plotted as a function of the distance away from the interface, allowing us to explore the dispersion of elements along the [001] growth direction of \LAO/BaO/\STO~ heterostructure. Intensity modulations are clearly evident in the Ti and Ba (inset of Figure \ref{Fig:3}(b)) line profiles, where the modulated peak positions represent the real stacking of the \TiOtwo~ and BaO layers, as indicated by the red and green arrows, respectively. This reveals a unique match between the simultaneously recorded HAADF and the EELS stack images, as evidenced in Figure S1. The irradiation damage induced by the electron source, although carefully minimized, inevitably results in a collapse of the near surface structure of \LAO~ (Figure S1(a)) and leads to the drop of La intensity in Figure \ref{Fig:3}(b). By checking the intensity modulation of Ba, one can identify the location of the BaO layers which appear to be confined within two atomic layers at the \LAO/\STO~ interface (inset of Figure \ref{Fig:3}(b) and Figure S1(b)). In addition, by using the intensity modulation as a reference, the width of the interface can also be estimated as to be three perovskite unit cells, shadowed in light blue of Figure \ref{Fig:3}(b). Considering the unavoidable broadening effect induced by shape of the electron probe and/or its tail\cite{kourkoutis2010atomic}, this value can be regarded as an upper limit of the chemical interdiffusion in the \LAO/BaO/\STO~ heterostructure.
\par
Wu \etal~ have shown thin film deposition using dissolved organic precursors \cite{wu2015atomic}. However, in their case TEM imaging indicated films which are far from having epitaxial quality in contrast to our perfect epitaxial monolayer.

\par
\begin{figure}
 \includegraphics[width=1\hsize]{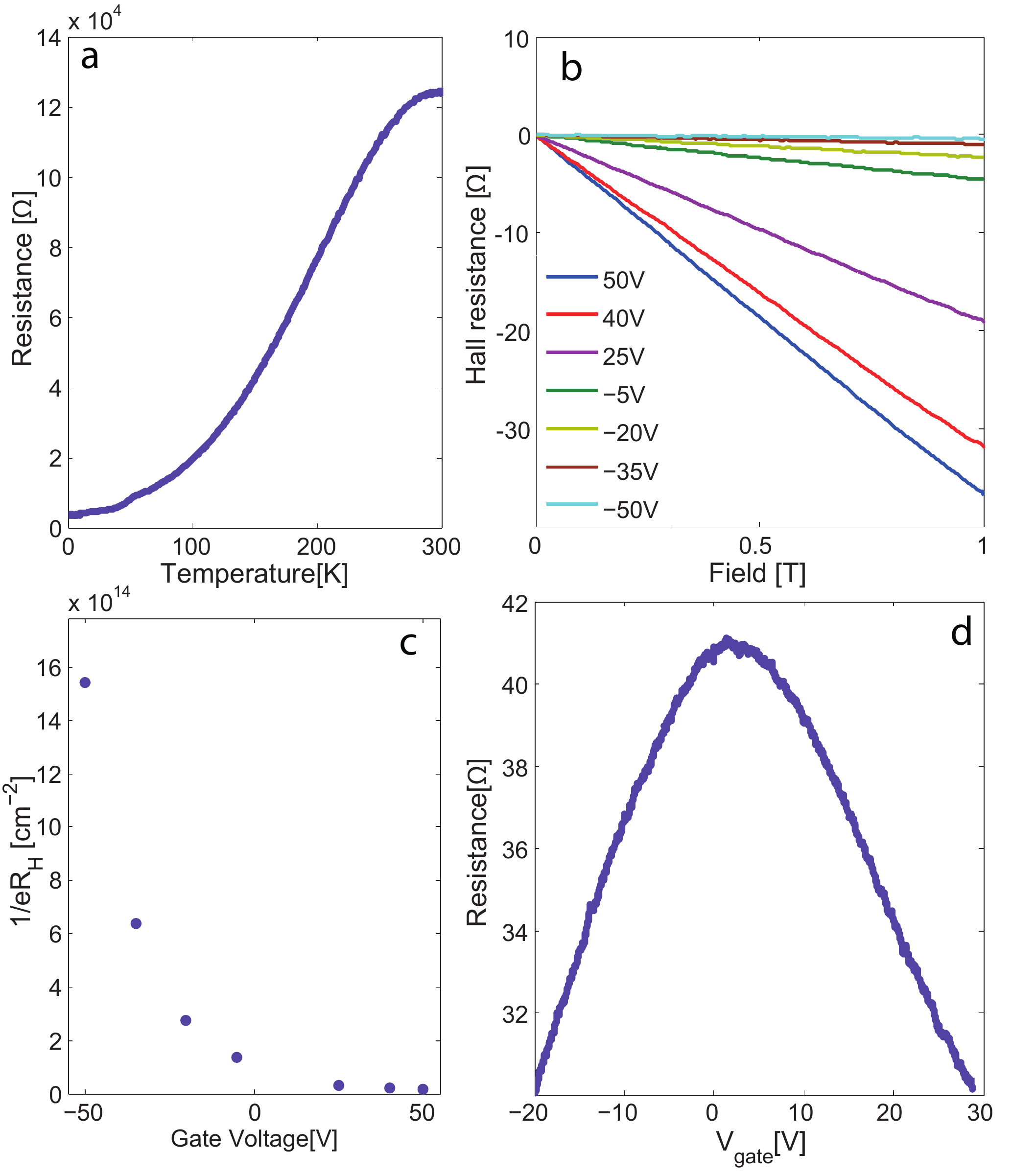}
  \caption {Transport properties of \STO/BaO\LAO~ interface.(a) Resistance as function of temperature (b) The Hall resistance measured as function of magnetic field for various gate voltages (c) Carrier density as function of gate voltage as calculated from the low field Hall coefficient. (d) Longitudinal resistance as function of gate voltage. A clear decrease in resistance is observed for increasing and decreasing of gate voltage.
}\label{Fig:4}
\end{figure}

In Figure \ref{Fig:4}(a) we show the sheet resistance \Rs~ of a typical sample measured as a function of temperature. As can be clearly observed \Rs~ is decreasing with temperature, i.e. the interface is metallic. In Figure \ref{Fig:4}(b) The Hall resistance as function of magnetic field is shown for various gate voltages. The slope of the Hall signal, the Hall coefficient $R_H$, is negative, corresponding to negative charge carriers. We plot $(eR_H)^{-1}$ versus gate voltage in Figure \ref{Fig:4}(c). In a naive, single band, picture the slope of the classical Hall effect is related to the charge carrier density by: $n = \frac{1}{eR_H}$. We note that the inferred density $n$ increases when applying negative gate voltage. This behavior is in contrast to what is expected from gating an electron doped sample. In Figure \ref{Fig:4}(d) we show \Rs~ as function of gate voltage. a clear non monotonic behavior is observed.

\par
The main evidence for the existence of both holes and electrons at the interface comes from the gate voltage dependence of $R_H$ and \Rs. As observed in Figure \ref{Fig:4}(c) $R_H^{-1}$ increases by orders of magnitude when applying negative gate voltages, in contrast to the expected behavior of n-type interface. Furthermore \Rs~ is nonmonotonic with gate voltage and has a clear maximum at $V_g=2$V.
\par
In the frame of bipolar transport of electrons with density $n$ and mobility $\mu_n$ and holes with density $p$ and mobility $\mu_p$ in the low field limit $R_H$ takes the form:
$R_H = \frac{p\mu^2_p-n\mu^2_n}{e(p\mu_p+n\mu_n)^2}$. When the condition $n\mu_n^2>p\mu_p^2$ is satisfied the Hall signal will be negative as observed in our measurements. Upon applying negative gate voltage $n$ decreases and $p$ increases. This results in effectively decreasing $|R_H|$. Furthermore, \Rs~ behaves nonmonotonically with gate voltage. In \STO/\LAO~ interfaces decreasing electron density results in lowering of electron mobility, \cite{bell2009dominant} which result in strong increase of the resistance. Therefore the decreasing resistance observed at V$_g<2$V can be explained only by accumulation of holes.
\par
A close inspection of our data suggests that $\mu_p>\mu_n$. For a symmetric electron-hole system the neutrality point should appear at the same gate voltage as the resistance maximum \cite{novoselov2004electric}. In our system the maximum in resistance appears where the sign of $R_H$ is negative, i.e. $p\mu^2_p<n\mu^2_n$. Assuming that the geometrical capacitance for both carriers is the same a decrease in $n$ is accompanied by an identical increase in p. Therefore, the decrease in resistance for decreasing gate voltage below $V_g=2$V must be in a regime where $\mu_p>\mu_n$ whereas the regime above $V_g=2V$ must be with $\mu_p<\mu_n$. We exclude other scenarios in the supplementary part.
\par
There are many mechanisms that can create n-type doping in \STO~ except for the polar scenario including: oxygen vacancies \cite{kalabukhov2007effect} and cation mixing \cite{willmott2007structural} close to the interface during film growth. However, non of these mechanisms can be responsible for hole doping. We therefore conclude that the polar structure of our samples is responsible for hole doping.
\par
It has been invoked that in n-type \LAO/\STO~ interface a hole layer is created on the \LAO~ surface due to charge transfer induced by the polar structure. Such a layer has not been observed and other explanations for the charge transfer have been proposed \cite{yu2014polarity}. In order to check if a mobile electron layer is created on the \LAO~ surface of our nominally p-type hetero-structure we fabricated a device with an additional Au electrode on the \LAO~ surface. If a mobile electron layer exists it will simply short the top electrode to the contacts. However, we merely observed a negligible leakage current, below our measurement capabilities, between the conducting interface and the top electrode. Finally, we note that while PLD deposited \LAO/SrO terminated \STO~ interface has been reported to be insulating our \LAO/BaO/\STO~ interface may be different because of either the preparation method or the variation in chemical composition (Ba instead of Sr).
\par
In summary, we developed a new solution monolayer epitaxy (SoME) method utilizing surface catalysis for depositing crystalline oxide layers. Our new SoME method does not require high temperature environments and it can be used instead of expensive physical deposition techniques or as an addition to them. The great variety of organic precursors make our method versatile and suitable for wide range of materials. To demonstrate our method we grow a BaO layer on top of a TiO$_2$ terminated \STO~. An additional \LAO~ layer is then deposited using PLD to realize a nominally p-type interface. We interpret the transport measurements of this new interface in the framework of bipolar conductivity where both electrons and mobile holes exist at the interface and can be controlled by gate voltage.

\section {Appendix: Experimental Methods }
\subsection{Solution monolayer epitaxy (SoME) of BaO monolayer}
In a typical preparation, inside a nitrogen-filled glovebox, about 10 mg of Ba$^{(II)}$ - isopropoxide (Sigma-Aldrich, 99.9\%) were weighed and subsequently suspended in 20 mL of anhydrous ethylene glycol (Sigma-Aldrich 99.8\%) in a glass vial. The vial was carefully sealed and taken out of the glovebox. It was then sonicated for 5 minutes in order to completely dissolve the barium isopropoxide in the ethylene glycol. The solution was then transferred to a round-bottom flask and a clean TiO$_2$-terminated \STO~ substrate was submersed in it. The flask was then purged three times with nitrogen and heated to $198^oC$, (reflux) under constant nitrogen flow. After 30 minutes the flask was let to cool back to room temperature and then opened. The substrate was then washed with acetone and isopropanol to remove the excess solution.
\subsection{Pulsed laser deposition and measurements}
We grow epitaxial layers of \LAO~ using reflection high energy electron diffraction monitored PLD on atomically flat BaO terminated $\{100\}$ \STO~ 0.5mm thick substrates in standard conditions, oxygen partial pressure of $1\cdot10^{-4}$ Torr and temperature of $780^oC$, as described in \cite{shalom2009anisotropic}.
Gold gate electrodes are evaporated to cover the back of the substrate. The leakage current is unmeasurably small ($<$1pA). Measurements were performed in a dilution refrigerator with a base temperature of 20mK at magnetic fields of up to 18 Tesla at various gate voltages at Tallahassee National High Magnetic Field Laboratory. Transport results were reproduced for 9, 10, 12 unit cells samples.
\subsection{Sample preparation for scanning transmission electron microscopy}
$<100>$ orientated cross-sectional lamellae were prepared using a focused ion beam milling FEI Helios NanoLab 400S dual-beam system. The lamellae were thinned by Ar ion milling at 2.5 kV in a Bal-Tec RES-101 system, followed by a 500 eV Ar ion milling to remove the damaged surface layers introduced by previous procedure with a Fischione NanoMill 1040 system.
\subsection{scanning transmission electron microscopy}
The atomic-resolution HAADF and EELS investigations were performed on an FEI Titan 80-300 microscope equipped with a spherical aberration corrector for the electron probe, running at 300 kV. In STEM mode, a probe size of about 0.1 nm and a semi-convergence angle of 25 mrad were utilized. For HAADF imaging, the inner collecting angle is 70 mrad. For EELS signal collection, the camera length parameter was set to a value of 48 mm, together with a 6.0 mm entrance aperture resulting in a semi-collection angle of about 40 mrad.

\section {Acknowledgements}
Special thanks to J-H Park, D. Graf for help in the magnet lab and to A. Gladkikh for the TOF-SIMS measurements. A.R and A.H equally contributed to this work. This work was supported in part by the Israeli Science Foundation under grant no.569/13 by the Ministry of Science and Technology under contract 3-11875 and by the US-Israel bi-national science foundation (BSF) under contract 2014047. A portion of this work was performed at the National High Magnetic Field Laboratory, which is supported by National Science Foundation Cooperative Agreement No. DMR-0654118, the State of Florida, and the U.S. Department of Energy. Doris Meertens of Research Centre J\"{u}lich is thanked for specimen preparation by focused ion beam milling. H. C. Du is thanked for helpful discussions.




\bibliographystyle{apsrev}

\end{document}